# Solar Wind at 33 AU: Setting Bounds on the Pluto Interaction for New Horizons

F. Bagenal[1], P.A. Delamere[2], H. A. Elliott[3], M.E. Hill[4], C.M. Lisse[4], D.J. McComas[3,7],
R.L McNutt, Jr.[4], J.D. Richardson[5], C.W. Smith[6], D.F. Strobel[8]

[1] University of Colorado, Boulder CO
[2] University of Alaska, Fairbanks AK
[3] Southwest Research Institute, San Antonio TX
[4] Applied Physics Laboratory, The Johns Hopkins University, Laurel MD
[5] Massachusetts Institute of Technology, Cambridge MA
[6] University of New Hampshire, Durham NH
[7] University of Texas at San Antonio, San Antonio TX
[8] The Johns Hopkins University, Baltimore MD

Corresponding author information:
    Fran Bagenal
    Professor of Astrophysical and Planetary Sciences
    Laboratory for Atmospheric and Space Physics
    UCB 600 University of Colorado
    3665 Discovery Drive
    Boulder CO 80303
    Tel. 303 492 2598
    bagenal@colorado.edu

**Abstract**

NASA's New Horizons spacecraft flies past Pluto on July 14, 2015, carrying two instruments that detect charged particles. Pluto has a tenuous, extended atmosphere that is escaping the planet's weak gravity. The interaction of the solar wind with Pluto's escaping atmosphere depends on solar wind conditions as well as the vertical structure of Pluto's atmosphere. We have analyzed Voyager 2 particles and fields measurements between 25 and 39 AU and present their statistical variations. We have adjusted these predictions to allow for the Sun's declining activity and solar wind output. We summarize the range of SW conditions that can be expected at 33 AU and survey the range of scales of interaction that New Horizons might experience. Model estimates for the solar wind stand-off distance vary from ~7 to ~1000 $R_P$ with our best estimate being around 40 $R_P$ (where we take Pluto's radius to be $R_P$=1184 km).

## 1 - Introduction

After a journey of over nine years the New Horizons spacecraft flies past Pluto on July 14, 2015. A scientific objective of the New Horizons mission is to quantify the rate at which atmospheric gases are escaping the planet [Stern 2008; Young et al., 2008]. At Pluto, the properties of the interaction of the escaping atmosphere with the solar wind depend not only on the rate at which the atmosphere is escaping from Pluto, but also vary with the solar wind conditions (e.g., flow, density, ram pressure, temperature, etc). Key to estimating Pluto's total atmospheric escape rate are measurements of the size of the solar wind interaction region.

The two New Horizons instruments that measure charged particles are the Solar Wind Around Pluto (SWAP) instrument [McComas et al., 2008b] and the Pluto Energetic Particle Spectrometer Science Investigation (PEPSSI) instrument [McNutt et al., 2008]. The SWAP and PEPSSI instruments (a) measure the deceleration of the solar wind from mass-loading by ionized atmospheric gases; (b) detect a shock upstream if the boundary to the interaction region is sufficiently abrupt; and (c) measure fluxes of Pluto ions when they are picked up by the solar wind. A specific challenge we face with New Horizons is that the spacecraft does not carry a magnetometer so that we will need to rely on monitoring of the interplanetary magnetic field (IMF) near Earth and propagating the field magnitude, field polarity (sector boundary) from measurements near 1 AU out to 33 AU.

New Horizons flies past Pluto at a distance of 32.9 AU from the Sun when the spacecraft is 31.9 AU from Earth with a one-way light time of 4h 25min. At the time of encounter, Pluto is 1.9° above the ecliptic plane on its eccentric orbit. New Horizons is heading nearly toward the nose of the heliosphere, in a direction toward the Galactic center. After the flyby there are options of targeting other objects in the Kuiper Belt within the following ~3 years at ~40 AU. Moving at 2.9 AU/year, the New Horizons spacecraft has enough fuel and communications capability to continue measuring the solar wind out to ~100 AU.

The Sun and the out-flowing solar wind vary on a wide range of timescales. The ~11-year solar cycle is associated with reversals of the polarity of the Sun's magnetic field. The solar cycle has a major impact on the coronal structure, which in turn drives the three-dimensional solar wind that fills and inflates the heliosphere. Around solar minimum, fast, steady wind arises at high latitudes from large circumpolar coronal holes and more variable, slow wind flows at lower latitudes from the streamer belt, coronal hole boundaries, and transient structures [e.g., McComas et al., 1998]. Around solar maximum, this simple structure breaks down with smaller coronal holes, streamers, and transients arising in the corona and leading to a complicated and highly variable solar wind structure at all heliolatitudes. The July 2015 New Horizons flyby of Pluto occurs as the Sun is in the descending phase of the solar cycle.

In section 2 of this paper we summarize the Voyager data obtained around the distance of Pluto's orbit. In section 3 we discuss the long-term, multi-decadal variability of the solar wind that needs to be taken into account when scaling the Voyager data to the New Horizons epoch. In section 4 we present examples of particle data obtained by New Horizons in the solar wind. In section 5 we make predictions for the scale of the region of the interaction of the solar wind with Pluto's atmosphere. We briefly discuss these modeling results and list our conclusions in section 6.

## 2 - Voyager Data

To survey solar wind conditions that the New Horizons spacecraft (and Pluto) could encounter in mid-2015 we have taken two approaches: (1) we have taken data obtained by Voyager 2 between 25 and 39 AU, and (2) looked at data from the New Horizons instruments on its trajectory to Pluto. We chose Voyager 2 because it has magnetic field data as well as reliable plasma. We took Voyager 2 data from 1988 to 1992 when it traversed from 25 to 39 AU at ecliptic latitudes of ~+4° to -8°. The maximum of solar cycle 22 was in ~1990, around the middle of our sample period. Voyager observations of the outer heliosphere are reviewed by Richardson et al. [1996a,b].

We show the trajectories of the two Voyagers and of New Horizons projected onto the ecliptic plane in Figure 1, highlighting the region of Voyager 2 data used (thick blue line). Figure 2 shows daily averages of solar wind flow speed (V), proton number density (n), proton temperature (T) and magnetic field magnitude (B) measured by Voyager 2 during this period. The solar wind flow is nearly radial. The average E-W angle from 20-40 AU is 0.75 degrees and the average N-S angle is 1.5 degrees. In a large interplanetary coronal mass ejection (ICME) at about 35 AU in 1991 these angles reached 5 degrees, but that was only in one very unusual event [Richardson et al., 1996c]. Densities and dynamic pressures are normalized by a $1/R^2$ factor to 32 AU. We note that at 32 AU pickup ions have a noticeable effect, decreasing the speed of the solar wind by several percent and increasing T, counteracting adiabatic cooling on expansion, resulting in a relatively flat temperature profile [Richardson et al., 2008].

The values of n, T and B show short-term (t~day-week) variations of a factor of 5-10 about a fairly steady average value. The solar wind speed shows smaller variations over days-weeks but shows a semi-periodic variation on a ~1.3-year timescale observed throughout the heliosphere [Richardson et al., 1994; Gazis et al., 1995]. These oscillations have not been reported in the current solar cycle. On a smaller scale, plasma changes were predominately due to Corotating Interaction Regions (CIRs), with speed increases roughly once per solar rotation. When Voyager 2 was near 32 AU (mid-1991) the solar wind speed jumped from 370 to 600 km/s and the density increased by more than a factor of ten. The effects of this Merged Interaction Region (MIR) persisted for about 30 days. Only a few shocks were observed per year at this distance by Voyager 2. More common are CIRs, which are typically observed once per solar rotation at this distance in the declining phase of the solar cycle [Lazarus et al., 1999]. Typical speed increases at a CIR at this distance are 30-50 km/s and these are often associated with energetic particle increases. The New Horizons spacecraft will take only a few hours to pass through the interaction region at Pluto, so we are hoping that the solar wind is relatively steady through this time.

We took the Voyager data shown in Figure 2 and made histograms of the parameters (Figure 3) and derived statistical quantities listed in Table 1. In Figure 3a the solar wind speed shows limited variation (±~10%) while the density distribution has a significant tail to higher densities. The derived quantities of solar proton flux (nV) and dynamic pressure ($P = \rho V^2 = n\, m_p\, V^2$ where $m_p$ is the mass of a proton) also exhibit significant tails (which is why we show median, 10$^{th}$ percentile, and 90$^{th}$ percentile values in Table 1 in addition to mean and standard deviation). Figure 3b shows distributions of magnetic field strength (B) and proton temperature (T) as well as derived quantities: Alfven Mach number ($M_{alf} = V/V_A$ where $V_A = B/(\mu_o \rho)^{1/2}$); and ratio of particle thermal pressure to magnetic pressure ($\beta =$

nkT/[B$^2$/2$\mu_\circ$]).  The high Mach number and low $\beta$ values indicate that at these distances the solar wind is a very cold, fast flow carrying the solar magnetic field.  Note that these Voyager plasma data do not include interstellar pick up ions which can make a significant contribution to the total thermal pressure at these distances, as discussed below.

Energetic particles in the vicinity of Pluto's distance in the solar wind were detected by the LECP instruments at Voyager 1 and Voyager 2 in the late 1980s and early 1990s.  Decker et al. [1995] reported on 28 keV – 3.5 MeV ions detected at Voyager 2 from 33 to 42 AU.  From 1991.5 to 1993.5 they observed energetic intensities that were recurrent with roughly the ~26-day solar rotation period, similar to the variations in the bulk plasma speed seen by the Voyager 2 PLS instrument.  There was also a long (~1 year) intensity increase associated with a pair of large travelling interplanetary shocks in 1991.  At shorter time scales, non-statistical variations of order 6 hours were also seen in the energetic particles.  Such multi-scale variations and correspondences between energetic particle and plasma measurements are typical in the solar wind, although structure tends to evolve at greater distances from the Sun.

To partially compensate for the lack of a magnetometer on New Horizons, we look for a relationship between solar wind ion properties and the local interplanetary magnetic field (IMF). We have examined several years of Voyager 2 data spanning both different heliocentric distances and solar wind conditions in the hope of resolving some reliable proxy measurement but have had only limited success.  We can characterize that effort by describing three different conditions.  First, the spacecraft observed typical solar minimum conditions in 1984 when it was at 14 to 15 AU from the Sun.  Second, when the spacecraft reached 32 AU in 1990 it was experiencing solar maximum conditions.  Third, when it saw declining phase conditions in 1993 the spacecraft was at 40 AU.  It may be possible to use observations such as these to infer a reasonable value for the unmeasured magnetic field once the solar wind conditions surrounding the encounter are known.  The fundamental problem with finding a proxy measurement for B is the merging of fast and slow streams along with ejecta that produces MIRs at solar maximum and CIRs at solar minimum. Fast wind and slow wind close to the Sun have established correlations between flow conditions and the average magnetic field, but merging modifies these properties.

As an example of what may be possible, Figure 4 shows the correlation between the Voyager 2 solar wind proton flux and the magnetic field intensity for the period Day Of Year (DOY) 90 to 220 of 1984.  This was solar minimum, or low solar activity levels. New Horizons encounter with Pluto during the declining phase after a weak solar maximum so we are expecting relatively low solar activity.  The correlation is helpful and can be refined once we better understand the encounter conditions.

## 3 - Long-Term Variations

Conventional wisdom has suggested that the basic solar wind output is primarily tied to the phase of the solar cycle, similar to the behavior of the three-dimensional structure.  However, McComas et al. [2008a] demonstrated that a much longer-term trend currently dominates over any smaller solar cycle effect. This multi-decade trend exhibits significantly reduced solar wind density, dynamic pressure and energy, and interplanetary magnetic field (IMF).

The most recent solar minimum (end of cycle 23) stretched into 2009 and was particularly deep and prolonged. In cycle 24 sunspot activity rose to a very small double peak. Meanwhile, the heliospheric current sheet developed large tilt angles, similar to prior solar maxima. McComas et al. [2013] recently extended their 2008 study and showed that the solar wind and IMF properties continued to drop through the prolonged solar minimum and maintained low values through the ~2012 "mini" solar maximum, illustrated in Figure 5. Figure 6 shows that the sunspot numbers went through a second maximum in early 2014 and that we expect New Horizons to be at Pluto as the Sun is on the declining phase of cycle 24.

From 2009 through DOY79 of 2013 the proton parameters are lower on average (compared to values typically observed from the mid-1970s through the mid-1990s) by the following factors [McComas et al. 2013]: density ~27%; temperature ~40%; solar wind speed and beta ~11%; thermal pressure ~55%; IMF magnitude ~31%, and radial component of the IMF ~38%; mass flux ~34%; dynamic pressure ~41%; energy flux ~48%; . With the proton dynamic pressure persisting near the lowest values measured in the space age (~1.4 nPa, compared to ~2.4 nPa typical from the mid-1970s to mid-1990s), these results have important implications for the solar wind's interaction with planetary atmospheres and magnetospheres.

The results of McComas et al. [2008a; 2013] indicate that the weak solar maximum and low solar wind output are driven by an internal trend in the Sun that is longer than the ~11-year solar cycle. For the purposes of this study, these results mean that the Voyager-based estimates from when Voyager 2 was around 30 AU (1988-1992) have to be scaled down significantly for predictions for New Horizons' flyby of Pluto in July 2015, as listed in Table 2.

Of particular importance for estimating the size of the interaction with an escaping atmosphere is the solar wind flux, nV. It is the solar wind momentum (strictly speaking $\rho$ V = n $m_p$ V) that is tapped to pick up ions that are created by photo-ionization and charge exchange of atmospheric molecules. We are particularly interested, therefore, in how the Voyager 2 values of nV scale down by a factor of (1-0.34)=0.66, giving a mean value scaled down from 3.24 to 2.14 km s$^{-1}$ cm$^{-3}$ (or 2.14 x 10$^9$ m$^{-2}$ s$^{-1}$). The median value scales down to 1.55 km s$^{-1}$ cm$^{-3}$ with 10$^{th}$/90$^{th}$ percentile values of 0.55/4.6 km s$^{-1}$ cm$^{-3}$ (Table 2).

New Horizons reaches 33 AU in July 2015 when the Sun is expected to be in the declining phase of an unusually low-activity solar maximum (Figure 6). Between 2012-2013 the monthly sunspot number exhibited an apparent peak at ~60, but late in 2013 and into the first half of 2014 the solar activity rose to peak sunspot numbers of ~100, declining through 2014-15, suggesting that the IMF intensity at 33 AU will be lower than experienced by Voyager at these distances.

## 4 - New Horizons Data

Since New Horizons flew past Jupiter in spring 2007 (for a gravity assisted en route to Pluto), the spacecraft has been in hibernation with annual periods of activity. The SWAP and PEPSSI instruments were initially turned off during hibernation but since 2012 have been taking data semi-continuously. By the time New Horizons encounters Pluto in July 2015 we will have three and a half years of solar wind data between 22 and 33 AU.

Figure 7 shows an energy-time spectrogram of SWAP data (20 eV/q to 8 keV/q) plus fluxes of 10s of keV to MeV particles measured by the PEPSSI instrument. We have picked a time that illustrates active behavior of the solar wind. The two panels show 168 days of data in the latter half of 2012 (DOY 189 to 357 where DOY=day-of-year) when New Horizons moved from 23.5 to 25 AU. The SWAP spectrogram shows the cold beam of solar wind protons (with a kinetic energy of ~1 keV) with lesser (few percent) fluxes of alpha particles, co-moving with the protons at twice the kinetic energy per charge. The ion energy distribution also shows a clear signal of interstellar pickup $H^+$ ions extending to four times the energy-per-charge of the proton beam [Randol et al., 2012, 2013].

## 5 - Predictions for Solar Wind Interaction with Pluto's Atmosphere

Having gathered a sense of the solar wind at Pluto's current distance from the Sun, we next address what is known about Pluto's escaping atmosphere and how the solar wind might interact with it.

### 5.1 Pluto's Atmosphere

Pluto's atmosphere was first detected in 1988 during stellar occultation [Elliot et al., 1989] and has since been determined to be primarily composed of $N_2$ with minor abundance of $CH_4$ and CO, with surface pressures of ~17 microbar [Young et al., 2001]. Pluto's low gravity implies that a significant flux of atmospheric neutrals can escape [Hunten and Watson, 1982; McNutt, 1989]. Estimates of escape rates range from as low as $1.5 \times 10^{25}$ $s^{-1}$ to as high as $2 \times 10^{28}$ $s^{-1}$ [Krasnopolsky 1999; Tian and Toon 2005]. Strobel [2008] estimates an outflow rate of $N_2$ of $2 \times 10^{27}$ $s^{-1}$ that is via hydrodynamic flow. The gas outflow velocities above the exobase estimated by Krasnopolsky [1999] and Tian and Toon [2005] are less than 100 m/s, from which Krasnopolsky [1999] concluded that Pluto could be an intermediate case between classic hydrodynamic escape and a static atmosphere.

In anticipation of observations from New Horizons, there have been several new studies of Pluto's atmosphere. Radiative-conductive-convective models by Zalucha et al. [2011a] and Zalucha et al. [2011b] for the lower atmosphere (troposphere and stratosphere) give higher lower boundary density ($3 \times 10^{13}$ $cm^{-3}$) and temperatures (121 K). Tucker et al. [2012] have modeled Pluto's atmosphere with a combined fluid/kinetic approach to calculate thermally driven escape of $N_2$ from Pluto's atmosphere. The fluid equations are applied to the dense part of the atmosphere while a kinetic (direct simulation Monte Carlo) approach is applied in the exobase region. The model found a highly extended atmosphere with an exobase at 6000 km at solar minimum with subsonic outflow and an escape rate comparable to the Jeans rates (i.e. enhanced Jeans escape). The lower boundary conditions for neutral density and temperature largely determine the overall profile of the atmosphere. The most recent atmospheric model of Zhu et al. [2014] indicates a denser and more expanded atmosphere with and escape rate of ~3.5 x $10^{27}$ $N_2$ $s^{-1}$ and an exobase at 8 $R_P$~9600 km.

We show in Table 3 four models (Strobel A, B, C, D) that show increasing escape flux and exobase height with larger methane abundance, and correspondingly lower outflow speed and density at the exobase height. Model A corresponds to the conditions for the atmosphere deduced by Lellouch et al [2015] and reflect the current best model. Detailed modeling by Volkov et al. [2011a,b] using a direct simulation Monte Carlo (DSMC) approach suggests that current understanding puts Pluto closer to Jean's escape rather than hydrodynamic regime. The details of thermal deposition and thermal conduction below the exobase tend to clamp the possible range of variability in the exobase altitude and other conditions there [Zhu et al., 2014]. The flow state can be characterized by the ratio $\lambda_c$ of the gravitational to thermal energy at the exobase (which we define as the radial distance at which the Knudsen number times $2^{1/2}$ ~1). With the ratio $\lambda_c$ ~5 for model 1, the conditions should be close to those of Jeans escape [Volkov et al. 2011], and the flow above can be approximated by free molecular flow (no collisions). This enables us to estimate conditions based upon the previous work of Chamberlain and others [Chamberlain 1963; Lemaire 1966; Chamberlain and Hunten 1987; Opik and Singer 1961; Aamodt and Case 1962] that were applied to the case of the Earth's geocorona [Bishop 1991].

In our models, the population of neutral molecules, here predominantly $N_2$, above the exobase can be divided into three components: (1) those with trapped ballistic orbits, i.e., originating effectively from the exobase but on orbits that fall back to the atmosphere; (2) hyperbolic escape orbits, i.e., having sufficiently large energies at their last collision to escape Pluto's gravitational field; (3) and those on "satellite orbits." The third population is energetically bound to Pluto, but on orbits whose periapses never go below the exobase. This population is somewhat of a contradiction since satellite orbits cannot be populated by atmospheric neutrals unless they suffer collisions above the exobase where, by definition, there are no collisions (infinite Knudsen number). A proper treatment would require a fully kinetic simulation in which such rare collisions would be taken into account.

Tucker et al. (2012) show that even rare collisions occurring above the exobase for Pluto affect the gas distribution and escape rates. Beth et al. (2014) show that "satellite particles" are non-negligible for H at Mars and for $H_2$ at high altitudes for Titan. We also note that Tucker et al. (2015) modeled Pluto's extended exosphere to examine the interaction with Charon. It was found that applying the free molecular flow approximation (collisionless) above the exobase was problematic because collisions above the exobase might produce a non-negligible population of "satellite particles". In their figure 2 it is shown that free molecular flow simulations applied above exobase underestimate the density profiles obtained by the fully collisional DSMC simulations.

The satellite population is limited by finite lifetime against ionization by charge exchange,

photoionization, or photo-dissociation, which can increase the energy above escape values. While the time scales for these processes are very long (~$10^9$ s or 30 years) in the vicinity of Pluto, the satellite orbits have unknown (possibly very long) lifetime. More importantly, the large variations in the solar wind flux conditions and solar luminosity can cause corresponding large changes in both ionization and exobase conditions, both of which can significantly affect this population. Furthermore, the presence of Pluto's moons on orbits out to ~55 $R_P$ could also disrupt and/or remove molecules on satellite orbits. If fully populated and stable, the population of molecules on satellite orbits would dominate the solar wind interaction region as discussed below, but for the reasons noted here, we expect this not to be the case. We think it is more likely that the escaping and ballistic (trapped) molecules populate the exosphere. Note that the satellite population is not present at comets, due to the low gravity of such small objects, and was not considered in the original treatment of the Pluto problem [Bagenal and McNutt 1989].

The escaping population contributes a declining density following an inverse square law with distance, as it must, far from Pluto. Close in, where the ballistic component dominates the density, the decline is more abrupt (~$r^{-5/2}$, cf. eqns, 45 – 47 of Chamberlain [1963] and surrounding text).

### 5.2 – Solar Wind – Atmosphere Interaction

As the atmosphere escapes Pluto's gravity and expands into space, the molecules are slowly ionized by solar UV photons and charge exchange with solar wind protons. The timescale for photoionization of $N_2$ is in the range of 1.2 to 3.3 x $10^9$ s depending on the UV activity of the Sun [Wegmann 1999]. This is 38-105 years, a significant fraction of the 248-year orbital period of Pluto. The charge-exchange rate [$H^+ + N_2 \longrightarrow H + N_2^+$] depends on the flux of solar wind protons and is nearly an order of magnitude lower than the photo-ionization rate [Wegmann 1999]. Thus, the timescale for removal of the neutral, escaping atmosphere is on the order of a few decades (by which time the escaping molecules have spread out thousands of Pluto radii) so that most of the atmosphere escapes into interplanetary space. Photo-dissociation of $N_2$ and dissociative ionization of $N_2$ also contribute to loss, but on even longer time scales [Huebner and Mukherjee 2015; Solmon and Qian 2005]. The small number of molecules that become ionized, however, can have a substantial effect on the solar wind. Once ionized, the molecule begins to gyrate around the ambient magnetic field and is immediately accelerated to the bulk speed of the solar wind. For a nominal solar wind speed of 380 km s$^{-1}$ the initial gyro-energies of $N_2^+$, $C^+$, $N^+$, and $O^+$ are 12-30 keV. For a typical ambient magnetic field of 0.1 nT the gyro-radii of pick up $N_2^+$ ions are 1.3 x $10^6$ km or 1000 >$R_P$. The momentum imparted to the picked up ion comes from the solar wind, which is correspondingly slowed down, stagnating the flow upstream of the planet, and potentially forming a shock [Galeev et al., 1985].

Initial studies of the solar wind interaction with Pluto's atmosphere (e.g. Bagenal and McNutt [1989]; Bagenal et al. [1997]) suggested that the solar wind interaction with Pluto's atmosphere would depend on whether the atmospheric escape flux is strong (producing a "comet-like" interaction) or weak (producing a "Venus-like" interaction). In both of these descriptions it is assumed that the planet's atmosphere/ionosphere and the solar wind could be considered as fluids. For many solar system bodies fluid descriptions of a plasma-obstacle interaction are often sufficient. Global-scale magnetohydrodynamic (MHD) models have been successful in capturing the basic structure of such plasma interactions. With the IMF being very weak at Pluto's orbital distance (Table 1), the length scales on which the plasma reacts are large compared with the size of the interaction region. For instance, at 33 AU the gyroradius of solar wind protons is ~23 $R_P$ and the pickup ion gyroradius is ~1000 $R_P$ [Bagenal and McNutt, 1989; Kecskemety and Cravens, 1993]. Furthermore, the upstream ion inertial length is comparable to the size of the obstacle (2-4 $R_P$) which could significantly alter the nature of the momentum transfer from the solar wind flow to the atmospheric ions.

The extended region of mass-loading far from Pluto makes it a "soft" obstacle to the supersonic solar wind. We contrast this with the expected "hard" obstacle created by a planetary magnetic field. However, inside the bow shock, in a region called the sheath, the plasma density increases as the decelerating flow approaches Pluto's presumably dense and bound atmosphere. The magnetic field, "frozen" to the flowing plasma, is compressed in the decelerating flow and correspondingly increases in strength. As some point, the pickup ions form a dense and "hard" obstacle to flow. In the cometary literature there is considerable confusion in the naming of the boundary (e.g., "cometopause", "ionopause", "collisionopause", "magnetic pile-up boundary", "contact surface") and its exact location (see discussions in Neugebauer [1990], Cravens [1991], and Gombosi et al. [1996]). We will adopt the term "interaction boundary" from here forward.

### 5.2.1 Cometary Model

The ions picked up in the unperturbed solar wind upstream of Pluto in an IMF of ~ 0.1 nT have large gyroradii (~600,000 km ~500 $R_P$). Predicting the location of the bow shock in the large ion gyroradius limit is difficult. Nevertheless, we compare the distance to the bow shock directly upstream of Pluto with predictions using the cometary models of Biermann et al., [1967] and Galeev et al., [1985] for partially mass loaded solar wind flow. In the fluid limit (or semi-kinetic if non-thermal pick-up ion pressure is included), pressure balance considerations provide the distance to the bow shock as the standard Galeev formula for solar wind stagnation point (normalized to Pluto's radius) as

$$R_s = Q_o\, m_i\, [4\pi\, V_{esc}\, \tau\, n_{sw} V_{sw}\, R_P]^{-1}\, [(\rho V)_c -1]^{-1} \qquad (1)$$

where $Q_o$ is the neutral escape rate, $V_{esc}$ is the neutral outflow velocity, $\tau$ is the ionization time constant, $n_{sw}$ and $V_{sw}$ are the unperturbed solar wind density and flow velocity and $(\rho V)_c$ is the critical loaded mass flux normalized to the upstream mass flux.

Biermann et al. [1967] showed that the solar wind flow is continuous until a value of $(\rho V)_c$ = 4/3 is reached (see also Flammer and Mendis [1991] for a more detailed treatment). This would make the factor $\zeta = [(\rho V)_c -1]^{-1} = 3$. From comparisons of models of comets (within the fluid regime) with models in the kinetic regime, Delamere (2009) argues that the kinetic case at Pluto is more consistent with $(\rho V)_c \sim 8/3$ giving $\zeta = 3/5 = 0.6$. In the fluid limit, each pickup ion in the mass-loaded upstream flow remains in the given fluid element. But in the large ion gyroradius limit, the pickup ion exits the solar wind fluid element laterally and the momentum transfer is far from complete on time scales less than the gyroperiod. In the hybrid simulations of Delamere [2009] the momentum transferred to the pickup ions in the upstream region is only a small fraction of the total pickup momentum. So the contamination to the flow is initially relatively small. As a result the bow shock moves closer to Pluto. With these limitations in mind, we apply two approaches to calculating the location of the interaction boundary: a cometary model where the gravity of Pluto is basically ignored above the exobase, and a coronal model where the gravity is included (more similar to the coronas of Mars and Earth).

For the cometary model we apply the Galeev formula and take the following nominal values: a neutral escape rate of $Q_o = 3 \times 10^{27}$ s$^{-1}$ [Zhu et al., 2014]; pickup ion mass of $m_i(N_2)$ = 28 amu; neutral escape speed $V_{esc}$ = 10 m/s; solar wind density $n_{sw} \sim 0.006$ cm$^{-3}$ = 6000 m$^{-3}$; solar wind speed $V_{sw} \sim 380$ km/s = $3.8 \times 10^5$ m/s; $R_P$=1184 km = $1.184 \times 10^6$ m. By plugging in these nominal values for photoionization timescale of $\tau \sim 1.5 \times 10^9$ sec, we get

$$Rs\ (R_{Pluto}) = 170\ R_{Pluto}\ \zeta \left(\frac{Q_o}{3 \times 10^{27}}\right) \left(\frac{10}{V_{esc}}\right) \left(\frac{1.5 \times 10^9}{\tau}\right) \left(\frac{0.006}{n(sw)}\right) \left(\frac{380}{V(sw)}\right) \quad (2)$$

Note that this formulation is only a rough approximation of where we could expect to see the stand-off distance of the solar wind upstream of Pluto. We aim to use New Horizons SWAP measurements of the upstream density $n_{sw}$ and speed $V_{sw}$ of the solar wind and its detection of the upstream boundary location (Rs) to estimate the net neutral escape rate ($Q_o$).

### 5.2.2 Adding Finite Gravity of Pluto

The situation at Pluto has the potential for being more complex due to the deviation of the density above the exobase from a strict inverse square law due to Pluto's gravitational field. If we re-examine Galeev's approach, it is convenient to define a "pick-up ion column density"

$$N_{pickup} \equiv \frac{\left[(\hat{\rho}\hat{u})_{crit} - 1\right] \tau \rho_\infty u_\infty}{m_{N_2}} \quad (3)$$

where $\tau$ is the total ionization time and all of the other quantities are as defined before with $N_2$ taken as the dominant pickup ion which slows the flow. A generalized stand-off equation can be written as

$$\int_{r_{standoff}}^{\infty} n(x)\ dx = N_{pickup} \quad (4)$$

For a simple inverse square density behavior we obtain the previous result for comets

$$N_{pickup} = \int_{R_s}^{\infty} n(x)\ dx = n_c r_c \left(\frac{r_c}{R_S}\right) = \frac{Q_0}{4\pi v_g} \frac{1}{R_S} \quad (5)$$

Note there is an implied relation in this case of

$$n_c r_c^2 \equiv \frac{Q_0}{4\pi v_g} \quad (6)$$

but this presupposes that the escaping component dominates both the ballistic and, potentially, satellite, component which is probably not the case with the finite gravity of

Pluto.

To provide stand-off estimates, we require the neutral density as a function of altitude above the exobase (assuming that the stand-off will occur above that level). For the limit in which λ ~ 0, the free molecular flow asymptotic limit results in a cubic equation with one real root in $\lambda^{1/2}$ if all the density components are included. However, the satellite component – if fully populated – dominates and we obtain

$$Rs(satellite) = \lambda_c r_c \left( \frac{8\lambda_c}{3\sqrt{\pi}} \frac{n_c r_c e^{-\lambda_c}}{N_{pickup}} \right) \tag{7}$$

For the numbers discussed here, this yields Rs ~1200 $R_P$, a distance from Pluto reached by New Horizons just over a day from closest approach for the nominal conditions. Even at such a large scale, the correction for ionization loss of the neutrals (the Haser correction in the cometary literature [Haser 1957], but also discussed in the appendix of Galeev et al. [1985] ) is not required. For bound molecules, the correction approach is more complicated in any event (see eqn. 121 in Chamberlain [1963] and the accompanying discussion).

The significant difference between the case of comets and the case of a gravitationally bound, evaporative exosphere is the presence of the ballistic component. For the conditions here, simply zeroing out the coefficient of the satellite component still leaves a cubic equation for $\lambda_{standoff}^{1/2}$, but the solution is no longer consistent with the asymptotic limit of the expression for the integrated column density. The asymptotic limit (with eqn. 95 of Chamberlain [1963]) yields a stand-off distance of ~26 $R_P$. The numerical value is closer to ~35 $R_P$ and the ballistic component only would yield ~ 30 $R_P$. New Horizons will reach 35 $R_P$ at ~50 minutes prior to closest approach to Pluto.

### 5.2.3 Numerical Simulation
To advance beyond a 1-D analytic description of the sub-solar stand-off distance, we need to turn to numerical simulations. Simulations of the solar wind interaction with comets is one of several instances where the fluid description is not appropriate and application of MHD is limited. The DS1/PEPE instrument measured a surprisingly asymmetric plasma environment near Comet Borrelly at 1.3 AU which Delamere [2006] attributed to large ion gyroradius effects. The pickup ion gyroradius in this case was comparable to the size of the interaction region, driving the plasma boundaries northward in the case of a northward-directed convection electric field. Delamere [2006] and Hansen et al. [2007] found that beyond 1.3 AU the fluid MHD approach is very limited.

While solar wind conditions at 30-50 AU dictate a kinetic treatment of all ion species, a hybrid approach is reasonable where the electrons are treated as a massless fluid, given that the electron inertial length is small (~50 km) compared to the gradient scale lengths of the extensive interaction region. To simulate conditions at Pluto we have applied the hybrid code first proposed by Harned [1982], and developed by Delamere et al. [1999]. The code assumes quasi-neutrality, and is non-radiative. We have developed a 3-D hybrid simulation for modeling the solar wind interaction with Pluto. Since our preliminary efforts [Delamere and Bagenal, 2004; Delamere, 2009], the code has been further developed to make significantly larger spatial domains feasible. Figure 8 shows three simulations of the solar wind (density n=0.006 cm$^{-3}$, flow speed V=380 km s$^{-1}$) interaction with an escaping atmosphere where Qo=3 x 10$^{27}$ molecules s$^{-1}$ and the outflow speed is taken to be 50, 25 and 10 m s$^{-1}$. The corresponding stand-off distances are 50, 90 and 170 R$_P$.

**6 – Discussion and Conclusions**
Figure 9 shows the location of this stagnation distance vs. solar wind flux (n$_{sw}$V$_{sw}$) for the various atmospheric escape models discussed in this paper for a range of values for Q$_o$ and V$_{esc}$. The median and 10$^{th}$ /90$^{th}$ percentile values of flux, scaled down from Voyager 2 values to the 2015 era, are shown as vertical lines. These models suggest that we can expect the New Horizons spacecraft to cross the upstream boundary anywhere from about 7 to ~1000 R$_P$. The models A-D apply the comet model shown in equation (2) to the Strobel atmospheric models listed in Table 3. Models E-H correspond to the same cases A-D but with V$_{exobase}$ = 100 m/s. The triangles are for the atmospheric models discussed above that include the effects of Pluto's gravity and correspond to exosphere populations of escaping (blue), ballistic (orange) and satellite (green) molecules. The green, orange and blue stars correspond to the interaction distances from the numerical simulations discussed above and presented in Figure 8.

Measurements from the SWAP and PEPSSI instruments on New Horizons will characterize the solar wind interaction with the ionized escaping atmosphere and quantify the mass-loading of the solar wind. Furthermore, comparison with these analytic and numerical models allow us to explore the sensitivity of the location of the interaction boundary to model conditions.

The main conclusions of this paper are:
1 – Voyager 2 measurements of the solar wind between 1988-1992, when scaled appropriately for the long-term weakening of the solar wind [McComas et al., 2008b; 2013], provide estimates of the plasma conditions in the solar wind that New Horizons can expect upstream of Pluto.

2 – When these scaled solar wind conditions are applied to simple (fluid) formulation of the distance at which we can expect the solar wind to be stagnated due to ionization and pickup of Pluto's escaping atmosphere, we find that stand-off distance could be anywhere from 7 to 1000 R$_P$ depending on (a) assumptions about the populations of neutral molecules in Pluto's exosphere and (b) the strength of the solar wind flux at the time of the flyby. Numerical simulations of the solar wind interaction produce similar estimates of the stand off distance. We estimate the likely stand-off distance to be around 40 R$_P$ (where we take Pluto's radius to be R$_P$=1184 km).

3 – We expect that the direction of the flux of recently picked up heavy ions will indicate the direction of the local interplanetary magnetic field. This IMF direction and ambient solar wind properties can then be compared with those measured in the inner heliosphere and propagated out to 33 AU.

*Acknowledgements*

The work at the University of Colorado was supported by NASA's New Horizons mission under contract 278985Q via NASW-02008 from the Southwest Research Institute (SwRI). Work at SwRI was supported as a part of the SWAP instrument effort on New Horizons under contract to NASA. All data shown in this paper are available via NASA's Planetary Data System. The model simulation output (Figure 8) is available by emailing Peter Delamere (Peter.Delamere@gi.alaska.edu).

FIG 1 – Trajectories of Voyagers 1, 2 and New Horizons through the outer solar system. The solar wind data used to predict conditions New Horizons will experience at Pluto were taken by Voyager 2 between 1988 and 1992 (thick blue line). The distances and speeds of these spacecraft are listed for mid-2015.

FIG 2 – Voyager 2 data obtained in the solar wind for years 1988 through 1992 when the spacecraft traversed from 25 to 39 AU.

FIG 3a - Histograms of solar wind properties based on Voyager2 data obtained in the solar wind for years 1988 through 1992 when the spacecraft traversed from 25 to 39 AU. (a) Bulk flow speed, (b) proton density, (c) proton flux, (d) dynamic pressure.

FIG 3b - Histograms of derived quantities based on Voyager2 data obtained in the solar wind for years 1988 through 1992 when the spacecraft traversed from 25 to 39 AU. (a) Magnetic field strength, (b) proton temperature, (c) Alfven Mach number, (d) ratio of particle thermal pressure to magnetic field pressure.

FIG 4 - Analysis of days 90 to 220 of 1984 Voyager 2 observations showing correlation between solar wind proton flux (y-axis) and |B| measured 2 days later (x-axis).

FIG 5: Top – Solar wind dynamic pressure in the ecliptic plane at ~1 AU, taken from IMP-8, Wind, and ACE and inter-calibrated through OMNI-2. Means (red), medians (blue), 25%-75% ranges (dark grey), and 5%-95% ranges (light grey) are shown time-averaged over complete solar rotations from 1974 through the first quarter of 2013. Bottom – shows the monthly (black) and smoothed (red) sunspot numbers and the current sheet tilt (blue) derived from the WSO radial model of Hoeksema et al., [1995]. From McComas et al. [2013].

FIG 6 – Solar Cyles 23 and 24 showing lower activity and predictions for New Horizons flyby of Pluto.

FIG 7 – Example of SWAP and PEPSSI observations during an active time (Jul. 7 – Dec. 22, 2012). The SWAP spectrogram shows the coincidence counting rate as a function of energy per charge and time for ions between 20 eV/q and 8 keV/q. The three PEPSSI traces show measured counting rates vs. time of protons between 5-7 keV (red, L11), 50-90 keV (blue, B01), and 220-370 keV (green, B04), where the channel L11 and B01 rates were scaled by a factor of 6 and 2, respectively, for display purposes. The 18-hour accumulations are sums across the total PEPSSI field of view.

FIG 8 – Numerical simulation (hybrid) of the solar wind interaction with Pluto's escaping atmosphere for 3 different cases of Vescape- 50, 25, 10 m/s. Based on Delamere [2009]. Solar wind conditions upstream are n=0.006 cm-3 and V=380 km/s.

FIG 9 – predictions of stand-off distance vs. solar wind flux. The vertical lines show median (solid) and 10/90-%-iles (dashed) values from Voyager 2 scaled to the New Horizons epoch. The models A-D apply the comet equation (2) to the atmospheric models listed in Table 3. Models E-H correspond to A-D but with Vexobase = 100 m/s. The triangles are for the models that include the effects of Pluto's gravity and correspond to exosphere populations of escaping (blue), ballistic (orange) and satellite (green) molecules. The green, orange and blue stars correspond to the interaction distances from the numerical simulations in Figure 8. New Horizons crosses the 1000 and 100 $R_{pluto}$ distances 23 and 1.3 hours respectively from closest approach to Pluto.

|  | 10% | Median | Mean Std Dev | 90% |
|---|---|---|---|---|
| $V_R$ [km/s] | 382 | 429 | 431 ± 40 | 482 |
| n @32AU [cm$^{-3}$] | 0.0020 | 0.0058 | 0.0070 ± 0.0053 | 0.0135 |
| nV [km s$^{-1}$ cm$^{-3}$] | 0.84 | 2.35 | 3.24 ±2.82 | 6.95 |
| T (K) | 3040 | 6650 | 8960 ± 8920 | 16800 |
| T (eV) | 0.26 | 0.57 | 0.77 ±0.77 | 1.45 |
| nkT [10$^{-4}$ pPa] | 1.18 | 5.32 | 9.64 ± 16.6 | 20.5 |
| P =$\rho V^2$ @32 AU [pPa] | 0.55 | 1.69 | 2.05 ± 1.64 | 3.98 |
| B [nT] | 0.08 | 0.15 | 0.17 ± 0.083 | 0.28 |
| $B^2/2\mu_o$ [10$^{-4}$ pPa] | 25.4 | 93.1 | 143 ± 152 | 311 |
| $V_{Alf}$ [km/s] | 21.7 | 45.0 | 54.9 ± 42.1 | 96.0 |
| $M_{Alf}$ | 4.6 | 9.5 | 11.3 ± 7.8 | 19.7 |
| Beta | 0.013 | 0.058 | 0.15 ±0.65 | 0.28 |

Table 1 – Statistical quantities derived from Voyager 2 plasma data between 25 and 39 AU (1988-1992).

|  | V2 Median | % Long-term change | V2 Scaled 10%-ile | V2 Scaled 90%-ile |
|---|---|---|---|---|
| $V_R$ [km/s] | 429 | -11% | 382 340 | 430 |
| n @32AU [$cm^{-3}$] | 0.0058 | -27% | 0.0042 0.0015 | 0.010 |
| T [K] | 6650 | -40% | 4000 1800 | 10,000 |
| nV [km/s $cm^{-3}$] | 2.35 | -34% | 1.55 0.55 | 4.6 |
| $P_{dyn}$ @32 AU [pP] | 1.69 | -41% | 1.00 0.32 | 2.3 |
| $P_{thermal}$ [$10^{-4}$ pP] | 5.32 | -55% | 2.4 0.53 | 9.2 |
| B (nT) | 0.15 | -31% | 0.10 0.05 | 0.19 |

Table 2 – Voyager 2 values of solar wind properties adjusted by approriate scaling factor to conditions during the New Horizons epoch. Note that the Pthermal does not include the interstellar pick ups which dominate the plasma thermal pressure at these distances.

| Model | Q0 (s-1) | CH4 fraction | p(μbar) | T (K) | V(exo) (m s-1) | n(exo) (cm-3) | Exobase (Rp) |
|---|---|---|---|---|---|---|---|
| Strobel A | 2.8E+27 | 0.0044 | 6.8E-09 | 69 | 4.6 | 7.1E+05 | 7.0 |
| Strobel B | 4.3E+27 | 0.0075 | 4.0E-09 | 59 | 6.5 | 4.9E+05 | 8.8 |
| Strobel C | 1.9E+27 | 0.0025 | 8.9E-09 | 76 | 3.5 | 8.5E+05 | 6.0 |
| Strobel D | 6.8E+26 | 0.0010 | 1.7E-08 | 90 | 1.6 | 1.3E+06 | 4.3 |

Table 3 – Atmospheric models adapted from the model of Zhu et al. [2014].

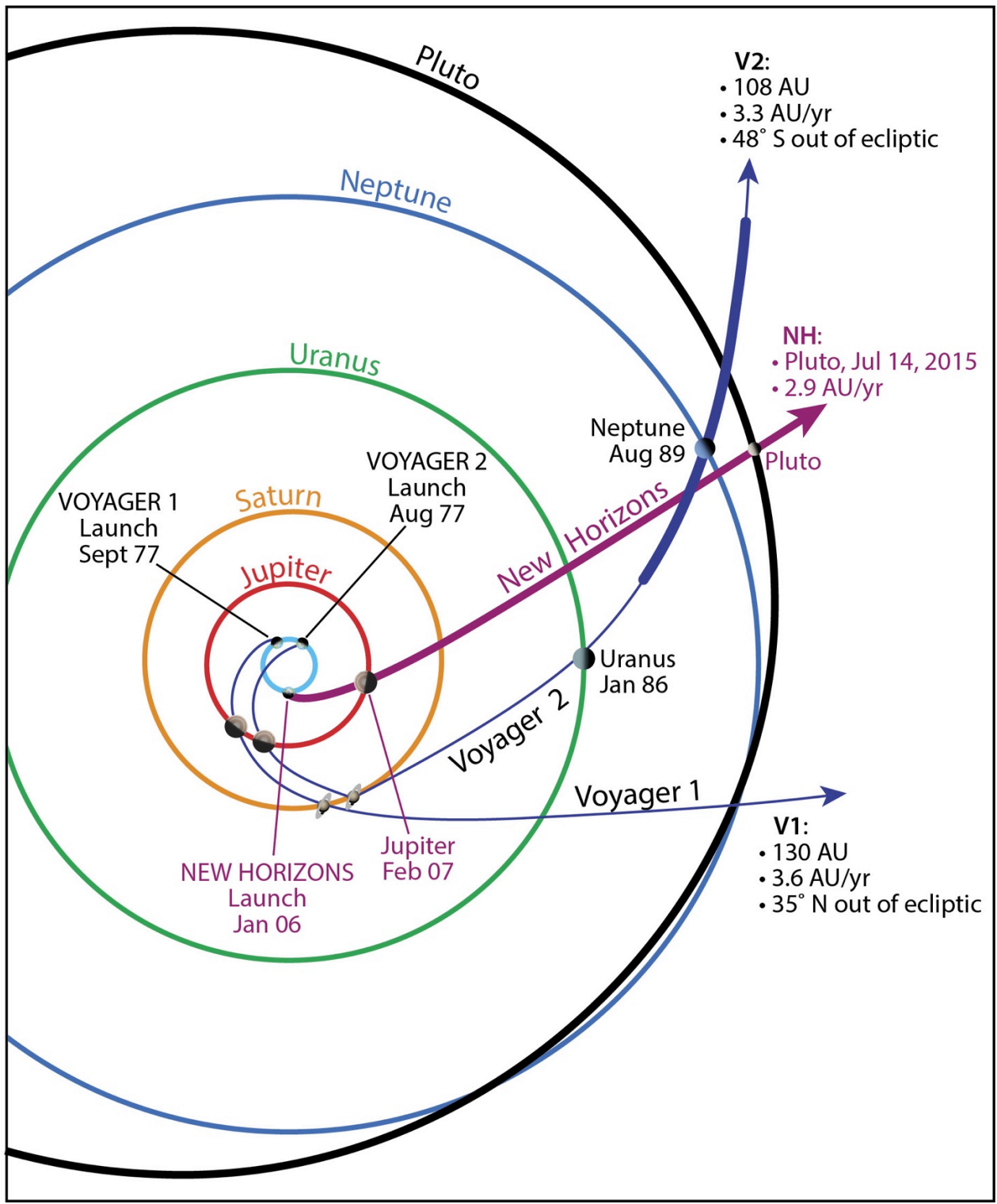

FIG 1 – Trajectories of Voyagers 1, 2 and New Horizons through the outer solar system. The solar wind data used to predict conditions New Horizons will experience at Pluto were taken by Voyager 2 between 1988 and 1992 (thick blue line). The distances and speeds of these spacecraft are listed for mid-2015.

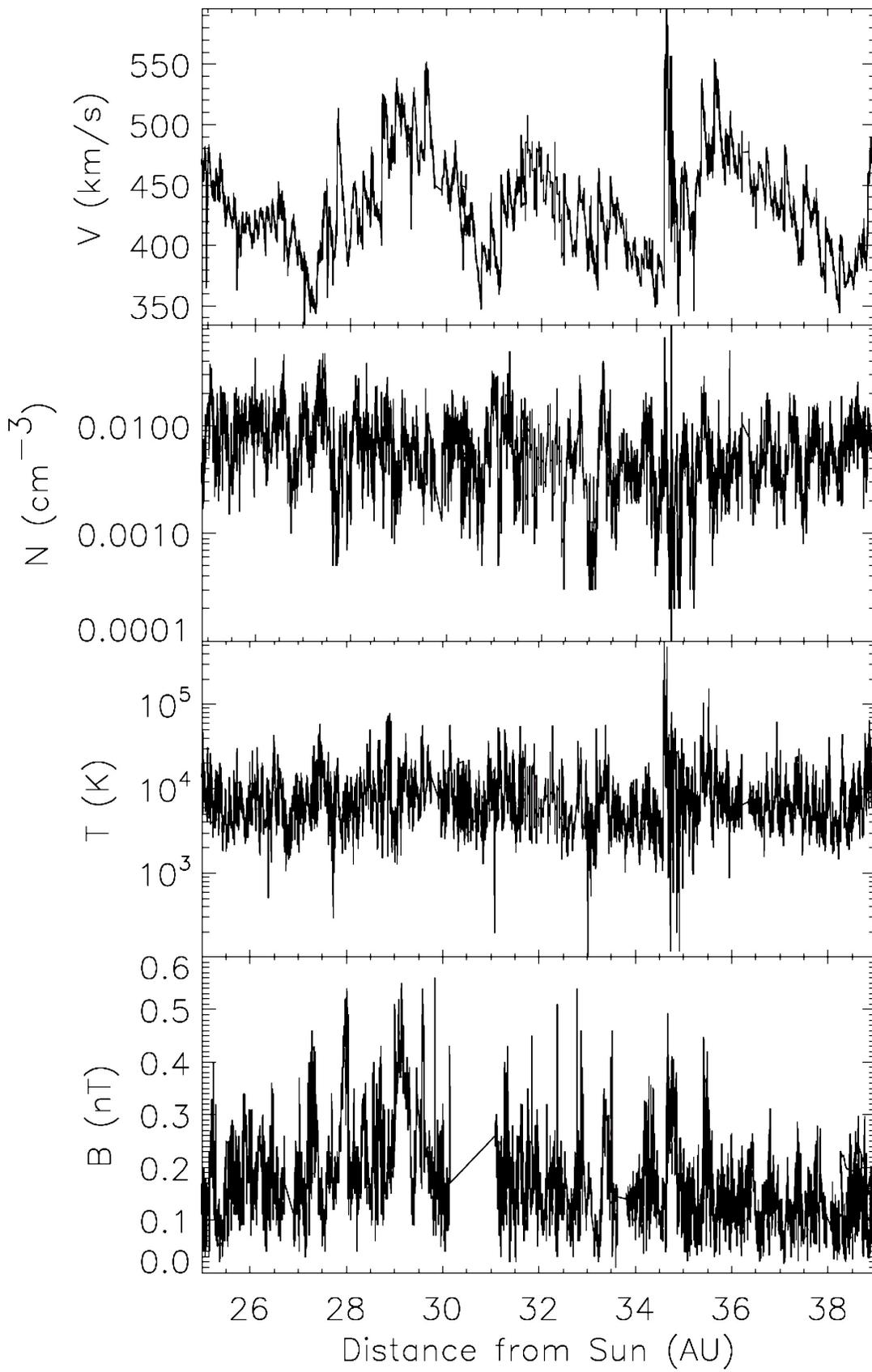

FIG 2 – Voyager 2 data obtained in the solar wind for years 1988 through 1992 when the spacecraft traversed from 25 to 39 AU. [2]

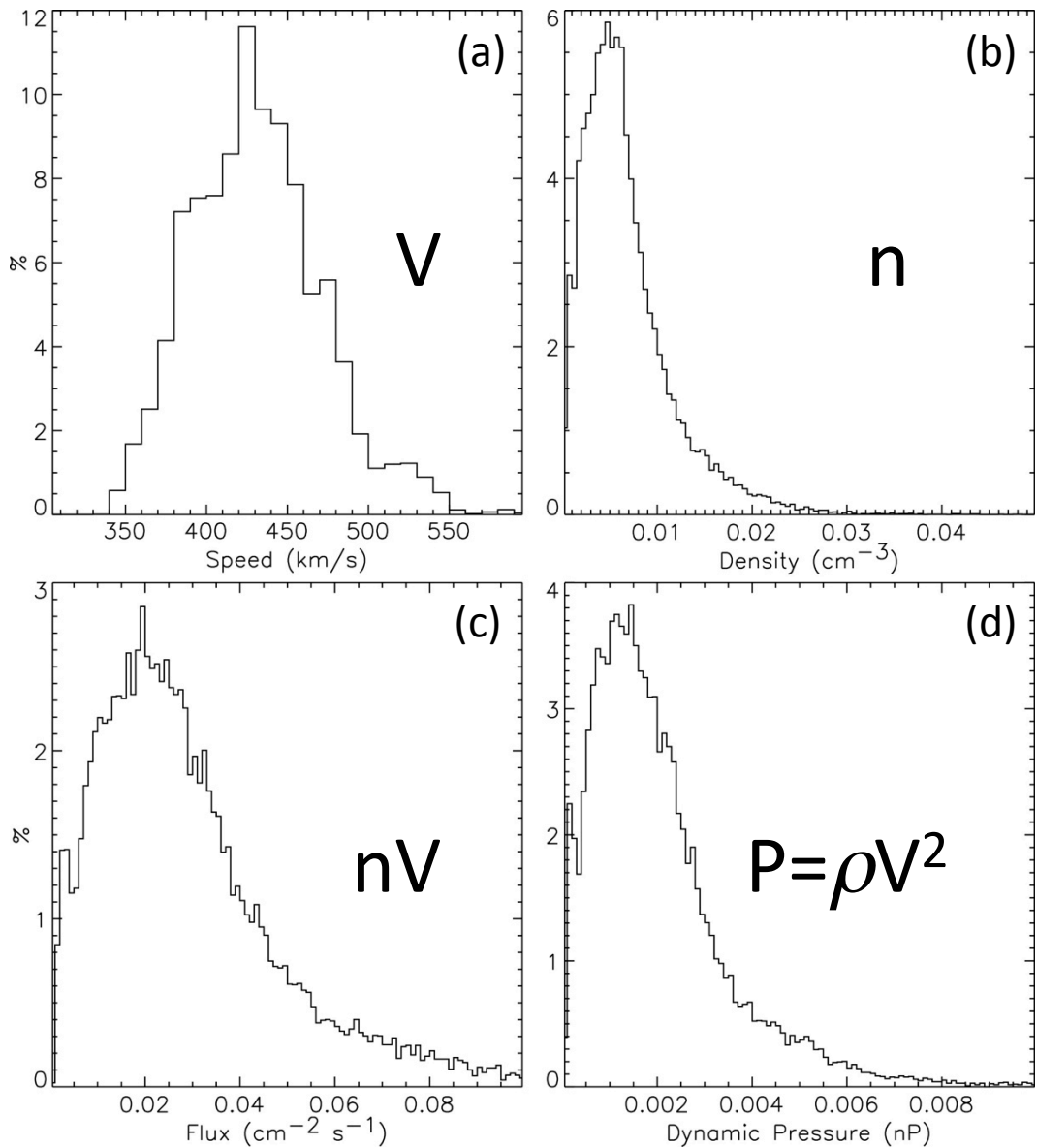

FIG 3a - Histograms of solar wind properties based on Voyager2 data obtained in the solar wind for years 1988 through 1992 when the spacecraft traversed from 25 to 39 AU. (a) Bulk flow speed, (b) proton density, (c) proton flux, (d) dynamic pressure.

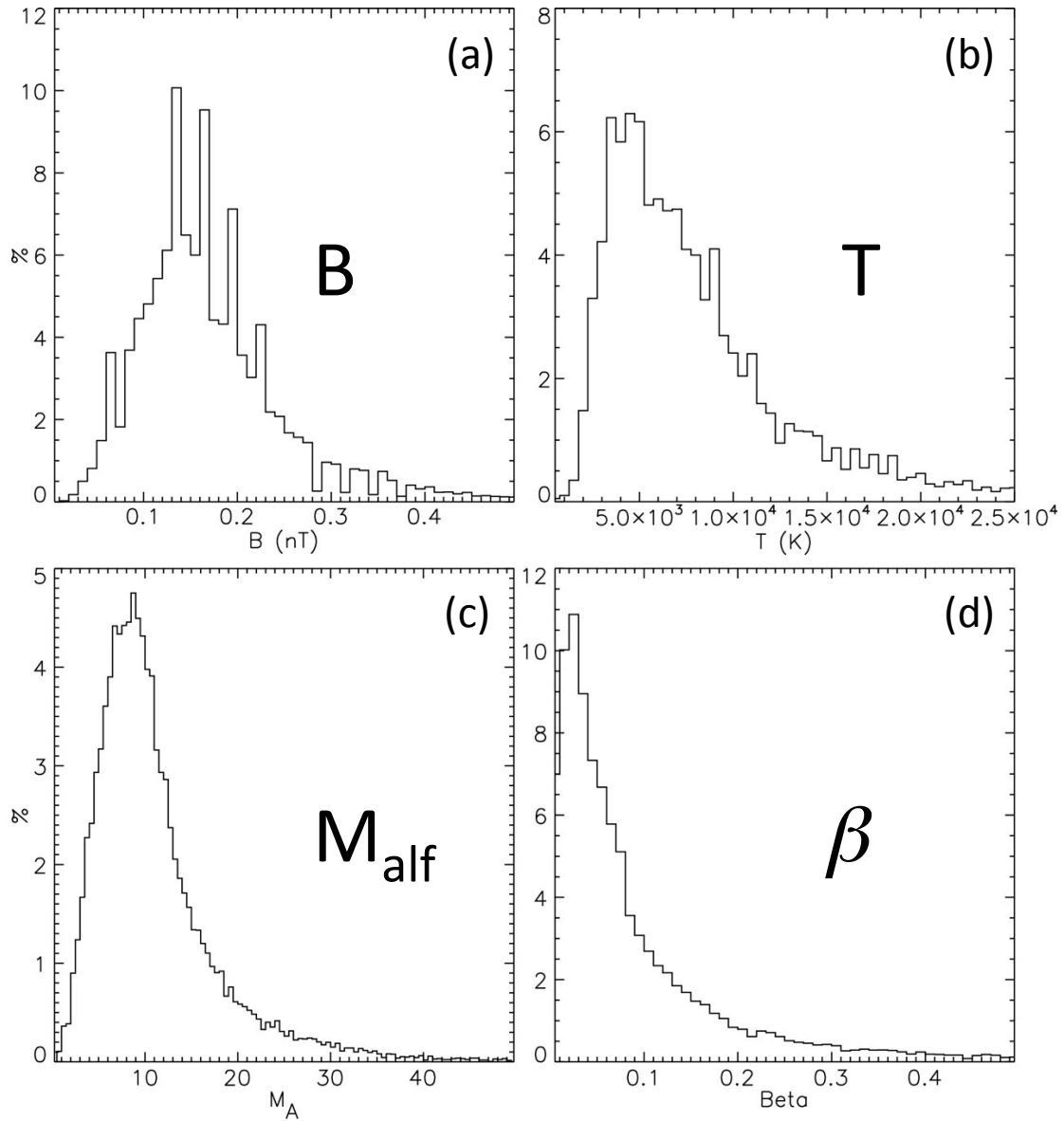

FIG 3b - Histograms of derived quantities based on Voyager2 data obtained in the solar wind for years 1988 through 1992 when the spacecraft traversed from 25 to 39 AU. (a) Magnetic field strength, (b) proton temperature, (c) Alfven Mach number, (d) ratio of particle thermal pressure to magnetic field pressure.

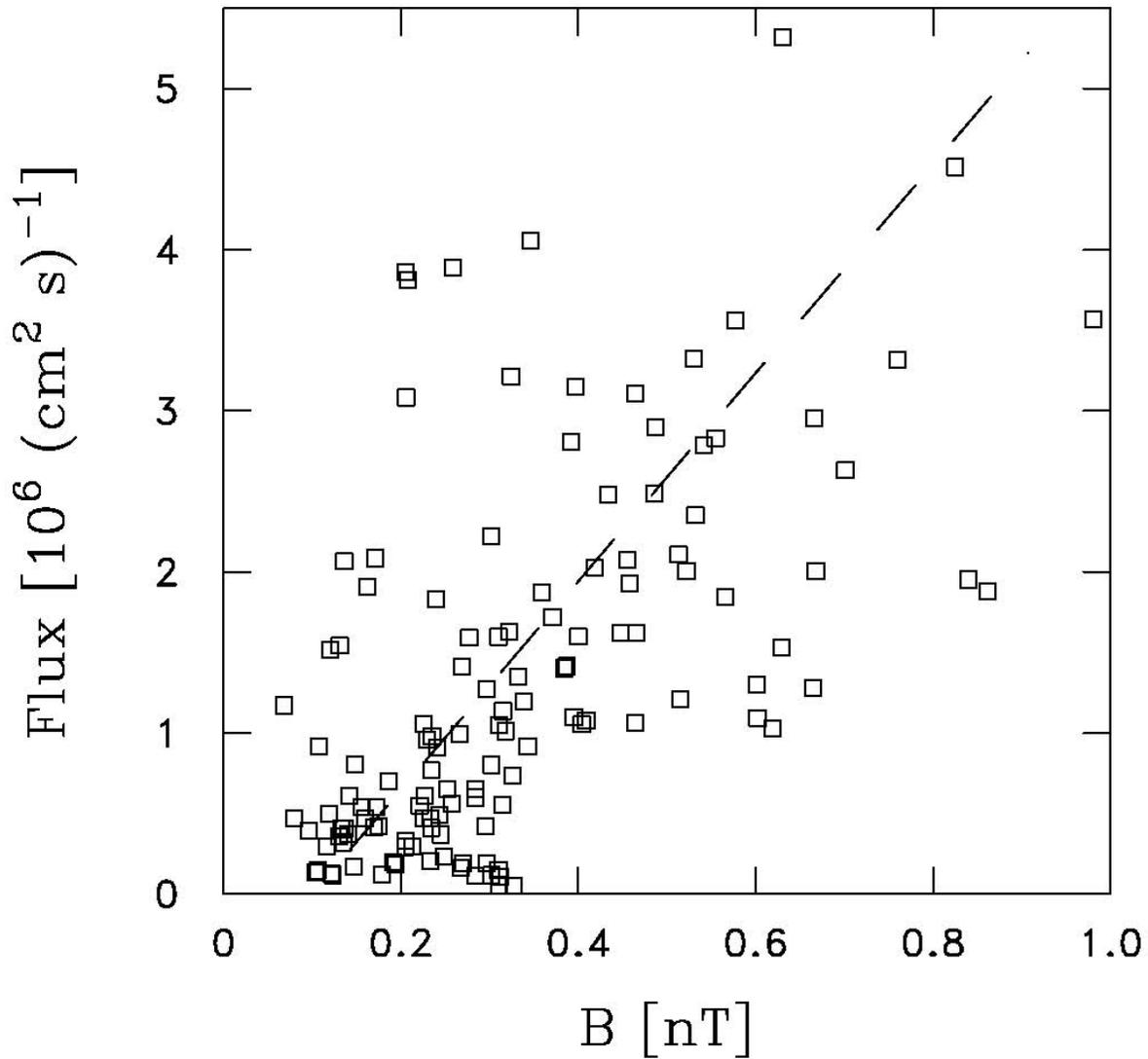

FIG 4 - Analysis of days 90 to 220 of 1984 Voyager 2 observations showing correlation between solar wind proton flux (y-axis) and |B| (x-axis).

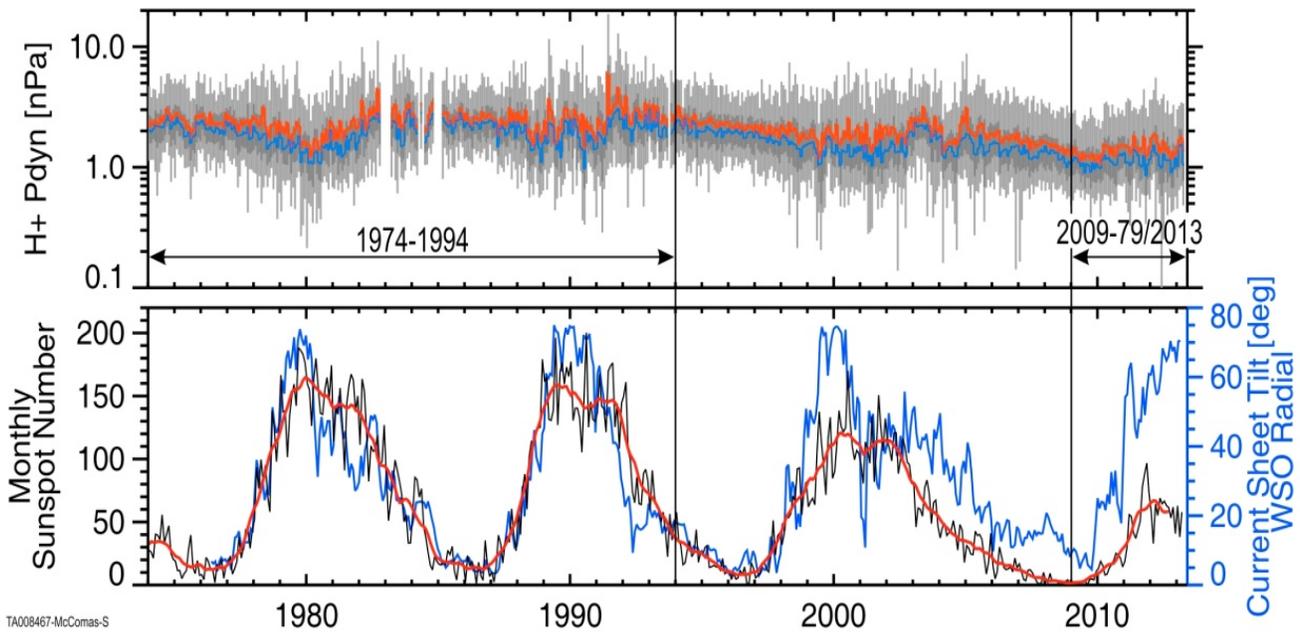

Figure 5: Top – Solar wind dynamic pressure in the ecliptic plane at ~1 AU, taken from IMP-8, Wind, and ACE and inter-calibrated through OMNI-2. Means (red), medians (blue), 25%-75% ranges (dark grey), and 5%-95% ranges (light grey) are shown time-averaged over complete solar rotations from 1974 through the first quarter of 2013. Bottom – shows the monthly (black) and smoothed (red) sunspot numbers and the current sheet tilt (blue) derived from the WSO radial model of Hoeksema et al., [1995]. From McComas et al. [2013].

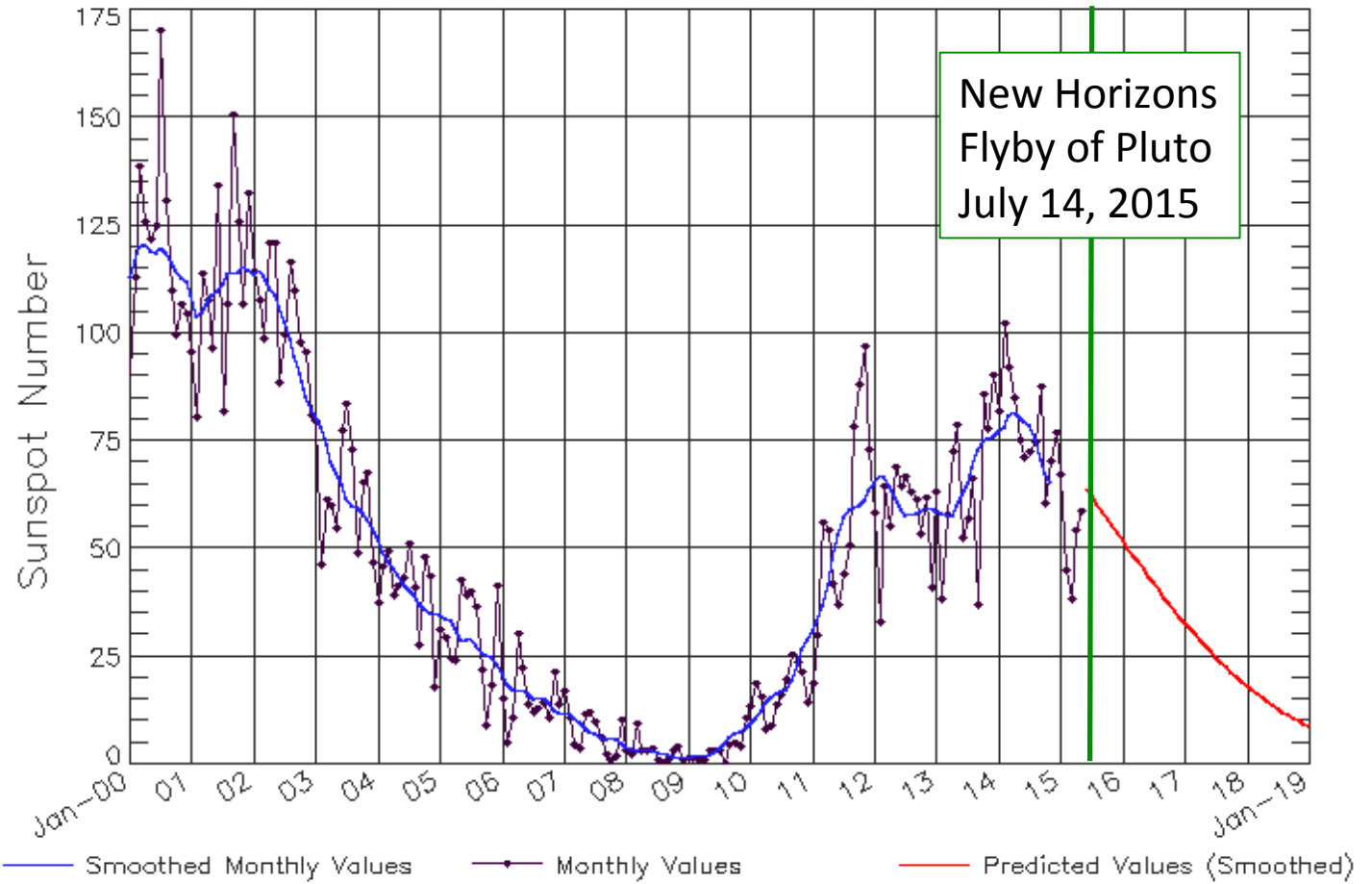

FIG 6 – Solar Cyles 23 and 24 showing lower activity and predictions for New Horizons flyby of Pluto.

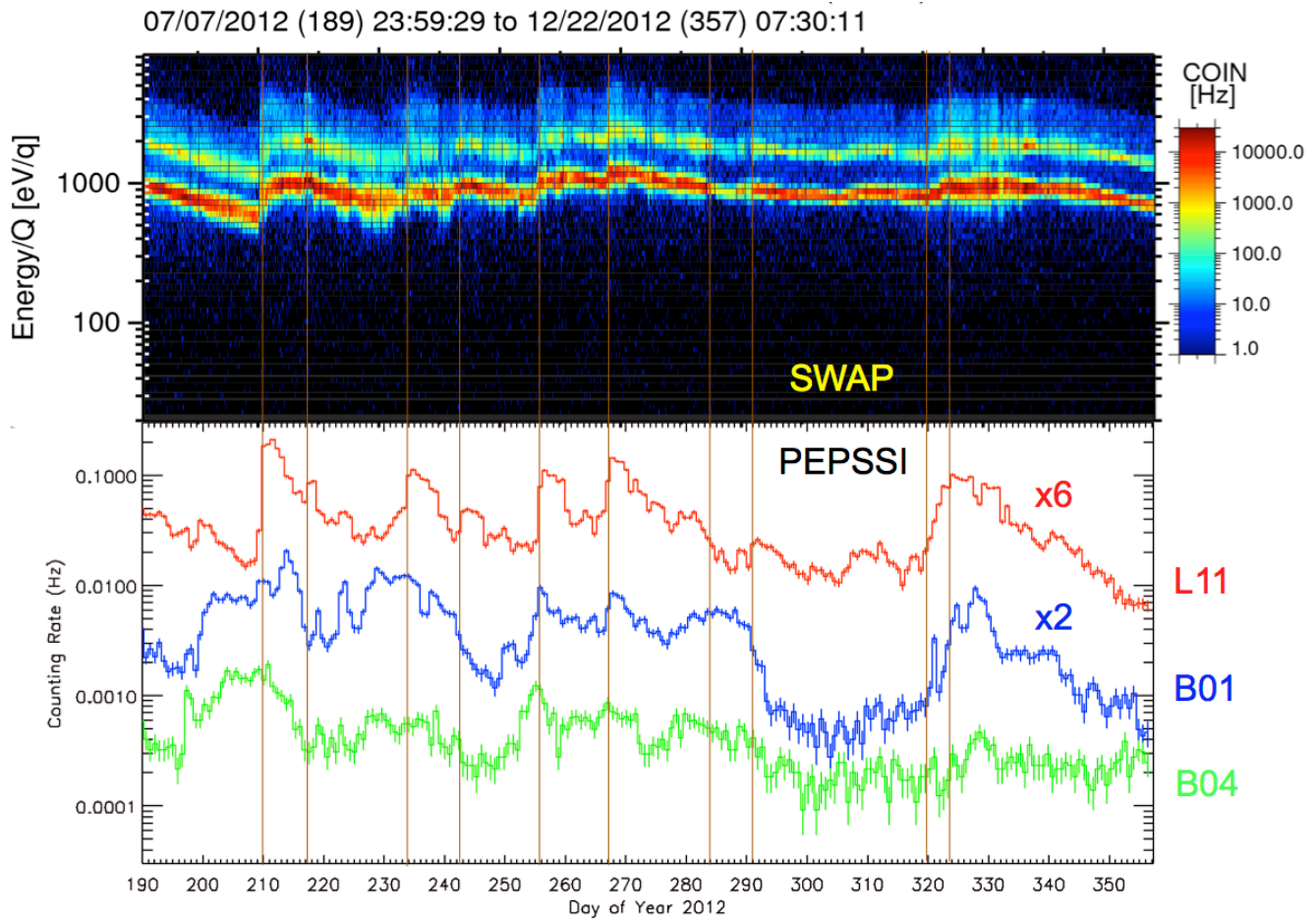

FIG 7 – Example of SWAP and PEPSSI observations during an active time (Jul. 7 – Dec. 22, 2012). The SWAP spectrogram shows the coincidence counting rate as a function of energy per charge and time for ions between 20 eV/q and 8 keV/q. The three PEPSSI traces show measured counting rates vs. time of protons between 5-7 keV (red, L11), 50-90 keV (blue, B01), and 220-370 keV (green, B04), where the channel L11 and B01 rates were scaled by a factor of 6 and 2, respectively, for display purposes. The 18-hour accumulations are sums across the total PEPSSI field of view.

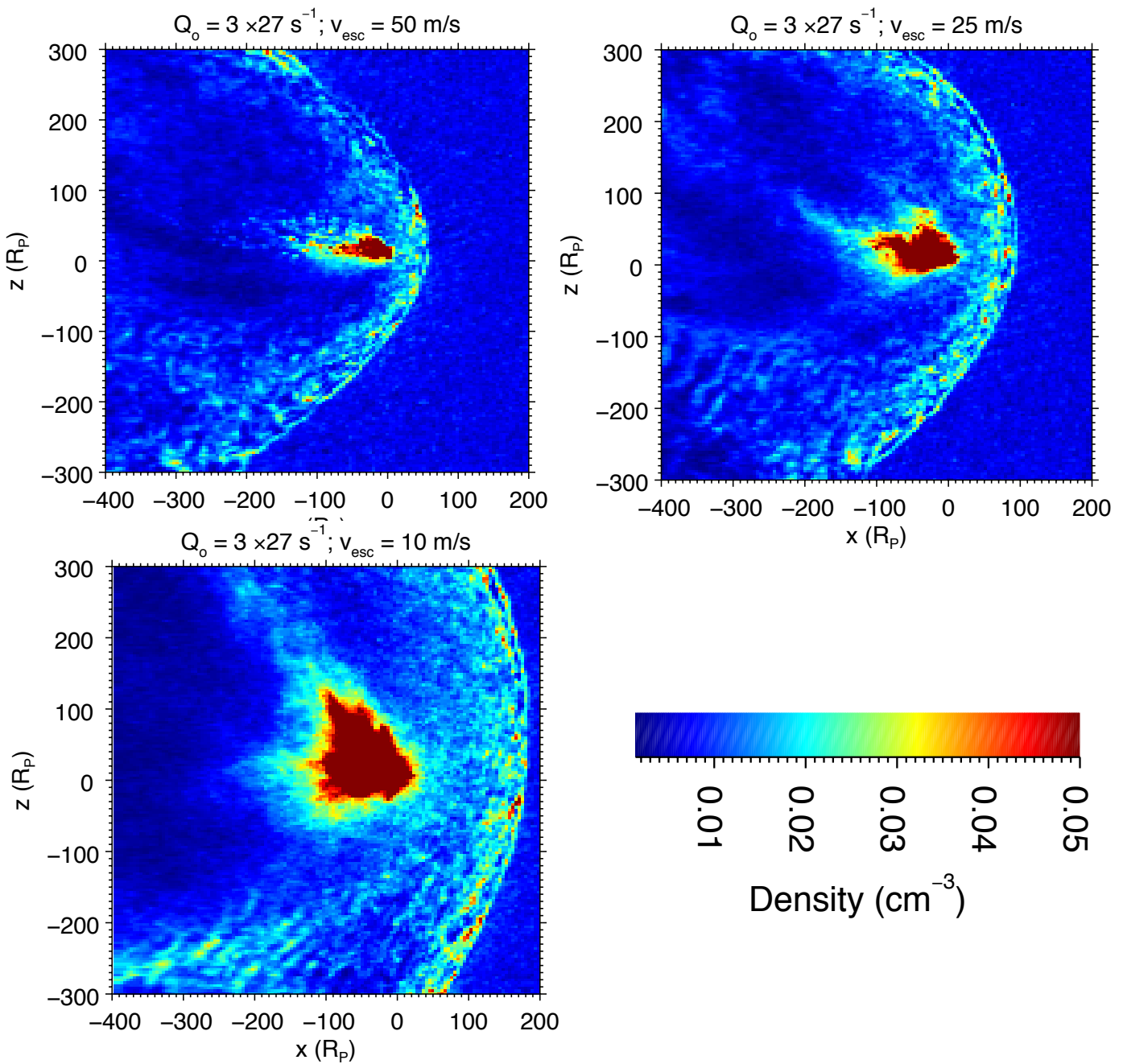

FIG 8 – Numerical simulation (hybrid) of the solar wind interaction with Pluto's escaping atmosphere for 3 different cases of Vescape- 50, 25, 10 m/s. Based on Delamere [2009]. Solar wind conditions upstream are n=0.006 cm-3 and V=380 km/s.

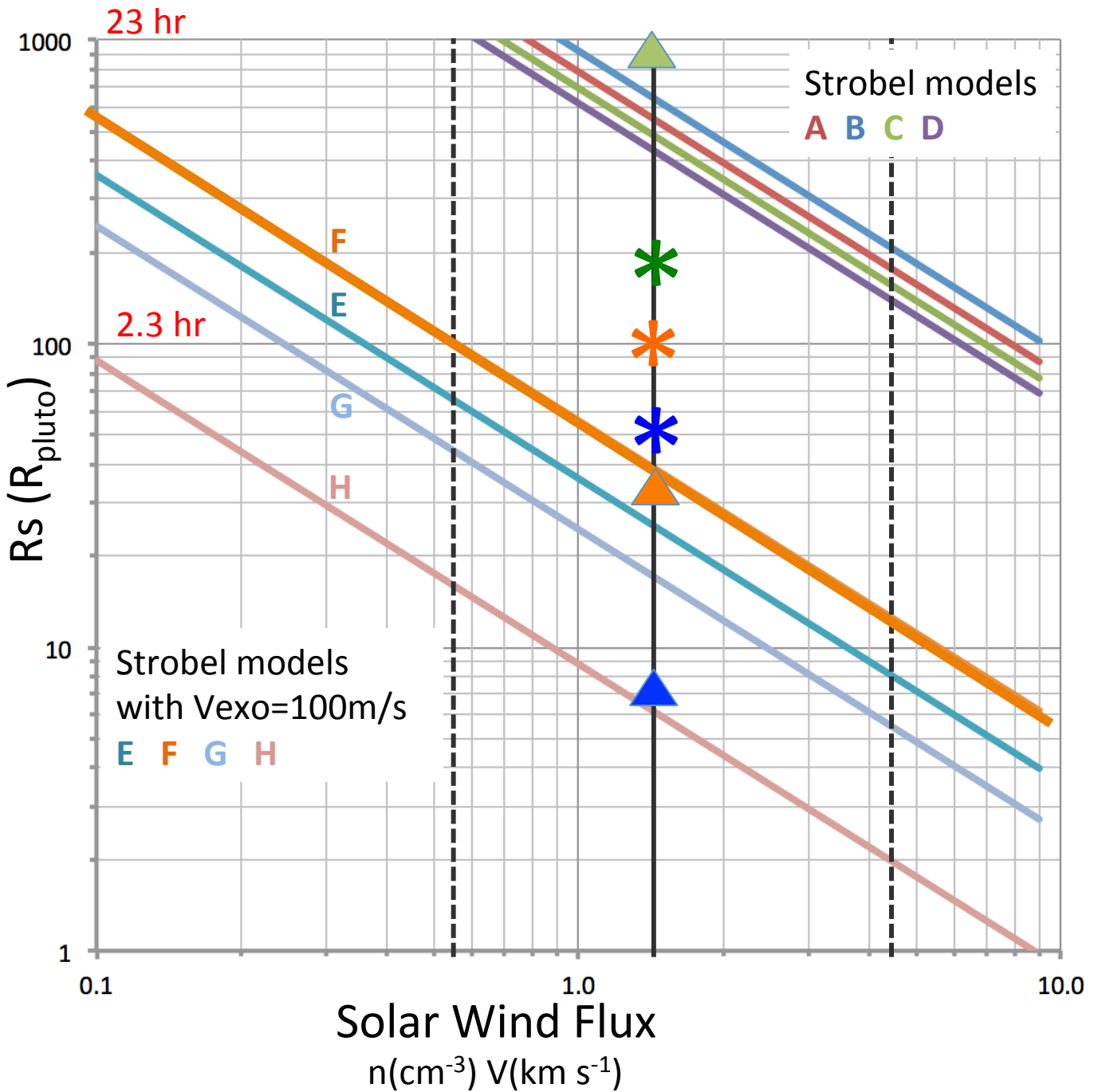

FIG 9 – predictions of stand-off distance vs. solar wind flux. The vertical lines show median (solid) and 10/90-%-iles (dashed) values from Voyager 2 scaled to the New Horizons epoch. The models A-D apply the comet equation (2) to the atmospheric models listed in Table 3. Models E-H correspond to A-D but with $V_{exobase}$ = 100 m/s. The triangles are for the models that include the effects of Pluto's gravity and correspond to exosphere populations of escaping (blue), ballistic (orange) and satellite (green) molecules. The green, orange and blue stars correspond to the interaction distances from the numerical simulations in Figure 8. New Horizons crosses the 1000 and 100 Rp distances 23 and 1.3 hours respectively from closest approach to Pluto.

10